\UseRawInputEncoding
\documentclass[a4paper,12pt]{article}
\pdfoutput=1
\usepackage{feynmp-auto,expdlist}
\usepackage{amsmath,amsfonts,amssymb}
\usepackage{graphicx}
\usepackage{enumerate}
\usepackage{hyperref}
\usepackage{latexsym}
\usepackage{hepnicenames}
\usepackage{enumerate}
\usepackage{soul}
\usepackage[normalem]{ulem}
\usepackage{wasysym}
\usepackage{makecell}
\usepackage{bbm}

\font\tenrsfs=rsfs10 at 12pt
\font\sevenrsfs=rsfs7
\font\fiversfs=rsfs5
\newfam\rsfsfam
\textfont\rsfsfam=\tenrsfs
\scriptfont\rsfsfam=\sevenrsfs
\scriptscriptfont\rsfsfam=\fiversfs

\numberwithin{equation}{section}

\usepackage{mathrsfs}
\usepackage{braket}
\usepackage{titling}
\usepackage{amsmath}
\usepackage{slashed}
\usepackage{amssymb}
\usepackage{epsfig}
\usepackage{graphicx}
\usepackage{color}
\usepackage{rotating}
\usepackage{hyperref}
\usepackage[table,xcdraw,dvipsnames]{xcolor}
\usepackage[compress,numbers,sort]{natbib}
\usepackage{colortbl}
\usepackage{pdflscape}
\usepackage{color}
\usepackage{mathtools}
\usepackage{colortbl}
\definecolor{Gray}{gray}{0.95}
\definecolor{RGray}{gray}{0.85}
\definecolor{CGray}{gray}{0.93}

\newcommand{\SU}{{\rm SU}}
\newcommand{\SO}{{\rm SO}}

\newcommand{\U}{{\rm U}}

\newcommand{\V}{{\cal V}}

\newcommand{\N}{{\cal N}}

\newcommand{\PQ}{{\rm PQ}}

\newcommand{\1}{{\textbf{1}}}
\newcommand{\3}{{\textbf{3}}}
\newcommand{\6}{{\overline{\textbf{6}}}}

\definecolor{nicered}{rgb}{0.7,0.1,0.1}
\definecolor{nicegreen}{rgb}{0.1,0.5,0.1}
\definecolor{red}{rgb}{1.0, 0, 0}
\definecolor{niceblue}{rgb}{0,0,0.8}
\definecolor{red}{rgb}{1.0, 0, 0}
\hypersetup{colorlinks,bookmarksopen,bookmarksnumbered,
linkcolor=blus,pdfstartview=FitH,urlcolor=rossos,citecolor=verde}
\allowdisplaybreaks

\definecolor{rosso}{cmyk}{0,1,1,0.4}
\definecolor{rossos}{cmyk}{0,1,1,0.55}
\definecolor{rossoc}{cmyk}{0,1,1,0.2}
\definecolor{blu}{cmyk}{1,1,0,0.3}
\definecolor{blus}{cmyk}{1,1,0,0.6}
\definecolor{bluc}{cmyk}{1,1,0,0.1}
\definecolor{verde}{cmyk}{0.92,0,0.59,0.25}
\definecolor{verdec}{cmyk}{0.92,0,0.59,0.15}
\definecolor{verdes}{cmyk}{0.92,0,0.59,0.4}

\def\eq#1{{Eq.~(\ref{#1})}}
\def\eqs#1#2{{Eqs.~(\ref{#1})--(\ref{#2})}}
\def\fig#1{{Fig.~\ref{#1}}}

\def\Table#1{{Table~\ref{#1}}}

\def\sect#1{{Sect.~\ref{#1}}}

\def\app#1{{App.~\ref{#1}}}

\def\vev#1{\left\langle #1\right\rangle}

\renewcommand{\bar}{\overline}

\newcommand{\beq}{\begin{equation}}
\newcommand{\eeq}{\end{equation}}
\newcommand{\bea}{\begin{eqnarray}}
\newcommand{\eea}{\end{eqnarray}}

\renewcommand{\(}{\left(}
\renewcommand{\)}{\right)}

\begin{document}



\begin{center}  
{\Large\Huge\bf\color{blus} 
Accidental SO(10) axion \\ from gauged flavour} \\
\vspace{1cm}
{\bf Luca Di Luzio} \\
\vspace{0.5cm}
{\it 
DESY, Notkestra\ss e 85, 
D-22607 Hamburg, Germany}\\
\vspace{0.5cm}
\begin{quote}
An accidental U(1) Peccei-Quinn (PQ) symmetry 
automatically arises in a class of SO(10) unified theories 
upon gauging the SU(3)$_{f}$ flavour group. 
The PQ symmetry is protected by the $\mathbb{Z}_4 \times \mathbb{Z}_3$ center 
of $\text{SO(10)} \times \text{SU(3)}_f$ up to effective operators of canonical dimension six. 
However, high-scale contributions to the axion potential posing a PQ quality problem 
arise only at $d=9$.
In the pre-inflationary PQ breaking scenario the axion mass window 
is predicted to be $m_a \in [7 \times 10^{-8}, 10^{-3}]$ eV, 
where the lower end is bounded by the seesaw scale 
and the upper end by iso-curvature fluctuations. 
A high-quality axion, that is immune to the PQ quality problem,  
is obtained for $m_a \gtrsim 0.02$ eV. 
We finally offer a general perspective on 
the PQ quality problem in grand unified theories. 
\end{quote}
\thispagestyle{empty}
\bigskip
\bigskip
\end{center}

\tableofcontents

\clearpage


\section{Introduction}
\label{sec:intro}

The Peccei-Quinn (PQ) \cite{Peccei:1977hh,Peccei:1977ur} 
mechanism relies on a global $\U(1)_{\rm PQ}$ 
symmetry that is broken in the infrared (IR) by the QCD anomaly.  
Global symmetries have no fundamental meaning, 
but they are rather understood to arise as accidental symmetries 
e.g.~in quantum field theories.  
Well-known examples are 
baryon and lepton number in the Standard Model (SM). 
Although an effective $\U(1)_{\rm PQ}$ is sometimes 
imposed ``by hand'', a proper \emph{PQ theory} should achieve that in an 
\emph{automatic} way. 
This was the term used in the early days \cite{Georgi:1981pu}, 
when first attempts were put forth to get an accidental $\U(1)_{\rm PQ}$ 
in grand unified theories (GUTs). 
The quick decline of the electroweak
Weinberg-Wilczek \cite{Weinberg:1977ma,Wilczek:1977pj} axion 
and the rise of the so-called \emph{invisible} axion 
\cite{Kim:1979if,Shifman:1979if,Zhitnitsky:1980tq,Dine:1981rt}, 
brought in another related puzzle, 
known as the \emph{PQ quality problem} \cite{Georgi:1981pu,Dine:1986bg,Barr:1992qq,Kamionkowski:1992mf,Holman:1992us,Ghigna:1992iv}: 
why is $\U(1)_{\rm PQ}$ an extremely good symmetry of ultraviolet (UV) physics? 
In fact, 
there is no reason 
to expect global symmetries to be exact, 
but   
even a tiny explicit breaking of $\U(1)_{\rm PQ}$ 
in the UV 
would spoil the PQ solution to the strong CP problem.  
Although the PQ quality problem is eventually a matter of UV 
physics which cannot be 
definitively assessed 
without a calculable theory of quantum gravity, the requirement of having the 
$\U(1)_{\rm PQ}$ to arise accidentally 
is a bare minimum 
that a sensible PQ theory in should structurally 
achieve.

It is the purpose of this work to revisit this question in 
the context of SO(10) GUTs. 
Since the SO(10) symmetry, by itself, is not sufficient 
to provide an automatic $\U(1)_{\rm PQ}$, 
some ingredient must be clearly added. 
An interesting possibility, put forth 
by Chang and Senjanovi\v c \cite{Chang:1987hz}, 
is to 
employ the $\SU(3)_f$ flavour symmetry of SO(10) 
(for related approaches without GUTs, see \cite{Chang:1984ip,Pal:1994ba}).
Remarkably, if the SO(10) representations are properly selected, 
the simultaneous presence of 
the horizontal ($\SU(3)_f$) and vertical 
(SO(10)) symmetry leads to an automatic PQ
symmetry. While Ref.~\cite{Chang:1987hz} focussed on the 
case of \emph{global} $\SU(3)_f$, 
motivated by the possibility of testing 
the $\SU(3)_f$ breaking dynamics at low-energy 
via the associated Goldstone bosons (familons), 
from a modern perspective it would be more satisfactory 
to get a $\U(1)_{\rm PQ}$ symmetry to arise from \emph{local} $\SU(3)_f$. 
The main obstacle for such a program is the cancellation of the 
$\SU(3)^3_f$ gauge anomaly, 
which requires to extend the fermion content of SO(10). 
In this work, we show how this can be consistently done, 
and we extend the analysis of \cite{Chang:1987hz} 
in several respects: $i)$ we point out that the same approach can be used beyond the 
renormalizable level, thus providing also a way to tackle the PQ quality problem; 
$ii)$ we identify the physical axion field and compute 
its low-energy couplings to 
SM matter fields. 
In this step, we realized that the charge assignment of \cite{Chang:1987hz} 
must be slightly modified, in order to avoid  
an alignment between 
two SO(10) Higgs representations
which would otherwise lead to a Weinberg-Wilczek axion;
$iii)$ we address $\SU(3)_f$ breaking dynamics. 

A relevant phenomenological feature 
of the accidental SO(10) axion is that its decay constant 
is bounded from above by the seesaw scale,  
which can be translated 
into a lower bound on the axion mass, $m_a \gtrsim 7 \times 10^{-8}$ eV.   
In fact, as we are going to show, 
in order to get an automatic PQ symmetry 
the SO(10) representations responsible for 
$\U(1)_{\rm PQ}$ breaking need to have a non-trivial SO(10) center. 
This immediately rules out the GUT-scale axion 
($m_a \lesssim 10^{-9}$ eV), in which 
the PQ breaking is connected to the first stage of SO(10) breaking, 
and selects instead SO(10) representations with $B-L$ breaking 
vacuum expectation values (VEVs)
at intermediate mass scales well below GUT. 
Other phenomenological constraints on the axion mass 
depend on whether the PQ is broken before or after inflation. 
In the former case, an upper bound of about $m_a \lesssim 10^{-3}$ eV 
originates from iso-curvature fluctuations generated by the massless axion field 
during inflation. On the other hand, a high-quality PQ symmetries requires 
a relatively heavy axion, 
$m_a \gtrsim 0.02$ eV,  
that is viable only if the PQ symmetry is broken after 
inflation and never restored afterwards. 
This scenario is threatened by a genuine domain wall problem, 
which however can be overcome by a small 
explicit breaking of the $\U(1)_{\rm PQ}$ (in terms of Planck-suppressed operators), 
compatibly with the PQ solution of the strong CP problem. 
We also discuss 
the relevance of astrophysical bounds in such case and 
the possibility that DM is wholly comprised by axions 
in the high-quality axion mass window. 

The paper is structured as follows. In \sect{sec:gaugingflav} 
we show how to obtain an accidental $\U(1)_{\rm PQ}$ symmetry 
upon gauging the flavour group of SO(10) and specify a minimal realistic model. 
In \sect{sec:PQquality} we extend the analysis beyond the renormalizable level, 
in order to identify at which operator level the $\U(1)_{\rm PQ}$ gets broken. 
Here, we also comment on the physical relevance of the 
PQ quality problem 
and on the possibility to address it within the present framework. 
\sect{sec:physaxion} is devoted to the identification of the 
physical axion field, which requires some care due to the fact that 
the axion gets kinetically mixed with neutral massive vectors 
arising from SO(10) breaking. Bringing the axion in a canonical 
form is actually a necessary step for computing its low-energy couplings to 
SM matter fields (as detailed in \app{sec:loweaxion}). 
Next, in \sect{sec:flavonsec} we address the cancellation of 
the $\SU(3)^3_f$ 
gauge anomaly as well as $\SU(3)_f$ breaking dynamics. 
In \sect{sec:axionexp} we discuss the phenomenological profile 
of the accidental SO(10) axion 
and conclude in \sect{sec:concl}, with a general perspective 
on the question of an automatic $\U(1)_{\rm PQ}$ in GUTs. 

\section{Gauging the way to the SO(10) axion}
\label{sec:gaugingflav}

A family of SM fermions plus a right-handed neutrino 
reside into a spinorial SO(10) representation, $\psi_{16}$, 
which gets triplicated in order to account for three SM chiral families.  
Schematically, a typical SO(10) Yukawa Lagrangian reads 
\beq
\label{eq:LYSO10}
\mathscr{L}_Y = 
y_{10} \, \psi_{16} \psi_{16} \phi_{10} 
+ \tilde y_{10} \, \psi_{16} \psi_{16} \phi^\star_{10} 
+ y_{\overline{126}} \, \psi_{16} \psi_{16} \phi_{\overline{126}} 
+ \text{h.c.} \, , 
\eeq
where we restricted 
for simplicity 
to a $10 + \bar{126}$ reducible Higgs representation, 
and we have taken a complex $\phi_{10}$ (as required by realistic fermion masses 
and mixings \cite{Babu:1992ia,Matsuda:2000zp,Matsuda:2001bg,Bajc:2005zf}), 
while the anti self-dual $\phi_{\overline{126}}$ is by construction 
complex.\footnote{The reader not familiar with SO(10) 
properties can find a basic introduction e.g.~in Sect.~2 of \cite{DiLuzio:2011mda}.}

In the $y_{10}, \tilde y_{10}, y_{\overline{126}} \to 0$ limit the global symmetry 
group of the SO(10) fermion sector 
comprising three copies of $\psi_{16}$ is 
\beq 
\label{eq:U3}
\U(3) = \U(1)_{\rm PQ} \times \SU(3)_f \, , 
\eeq
where the abelian factor is a PQ symmetry, since it is anomalous under QCD. 
In fact, the chiral embedding of SM matter into the spinorial of SO(10) 
implies a non-zero $\U(1)_{\rm PQ}$-$\SU(3)_c^2$ anomaly (cf.~\sect{sec:physaxion}).

In the following, we wish to argue 
that the \emph{gauging} of $\SU(3)_f$ leads to an accidental $\U(1)_{\rm PQ}$ 
in the full Lagrangian, if the SO(10) Higgs representations are properly chosen 
along the lines of \cite{Chang:1987hz}.   
Since the SO(10) spinors transform in the fundamental of $\SU(3)_f$, 
$\psi_{16} \sim \textbf{3}$, 
in order to make the Yukawa Lagrangian $\SU(3)_f$ invariant 
we 
need to 
assign both 
$\phi_{10} \sim \overline{\textbf{6}}$ 
and $\phi_{\overline{126}} \sim \overline{\textbf{6}}$,  
since SO(10) contractions are 
symmetric.\footnote{On the contrary, a $\phi_{120}$ would have to be assigned 
to a $\textbf{3}$, 
being the $\psi_{16} \psi_{16} \phi_{120}$ SO(10) contractions
antisymmetric.}   
Then the $\tilde y_{10}$ term in \eq{eq:LYSO10} is forbidden 
by $\SU(3)_f$ gauge invariance and a $\U(1)_{\rm PQ}$ automatically arises in the Yukawa sector:   
\beq 
\label{eq:PQLY}
\psi_{16} \to e^{i\alpha} \psi_{16} \, , \quad
\phi_{10, \, \overline{126}} \to e^{-2i\alpha} \phi_{10, \, \overline{126}} \, , 
\eeq
where $\alpha$ indicates the parameter of the $\U(1)_{\rm PQ}$ 
transformation and PQ charges are explicitly denoted as 
$\PQ(\psi_{16}) = 1$, $\PQ(\phi_{10,\overline{126}}) = -2$, etc.

Other Higgs representations need to be introduced for the
spontaneous symmetry breaking of $\SO(10) \times \U(1)_{\rm PQ}$ 
($\SU(3)_f$ breaking will be discussed separately in \sect{sec:SU3breaking}). 
To this end we consider a (real) adjoint $\phi_{45}$ and a $\phi_{16}$. 
Their role is the following:
\begin{itemize}
\item $\phi_{45}$: this is the smallest representation which can break 
$\SO(10)$ down to a rank-5 
sub-group featuring an unbroken $\U(1)_{R} \times \U(1)_{B-L}$ 
Cartan subalgebra, although it requires to go beyond the tree approximation 
for the minimization of the scalar potential 
\cite{Bertolini:2009es,Bertolini:2012im,Bertolini:2013vta,Graf:2016znk}. 
As far as concerns the transformation properties of $\phi_{45}$ under $\SU(3)_f$, 
since it does not couple to $\psi_{16}$ at the renormalizable level, 
it can be taken to be a singlet. 
In fact, even if we were to assign $\phi_{45}$ to a 
non-trivial $\SU(3)_f$ representation (say a fundamental), 
a PQ breaking 
operator of the type $\phi_{45}^3 \equiv 
\epsilon_{(3)}^{abc} 
(\phi_{45})_{a}^{ij}
(\phi_{45})_{b}^{jk}
(\phi_{45})_{c}^{ki}$ 
would be always allowed by $\SO(10) \times \SU(3)_f$ invariance. 
Similar considerations apply to other SO(10) representations 
with a trivial SO(10) center, like $54$ and $210$, 
which could be similarly employed in place of $\phi_{45}$. 

\item $\phi_{16}$: the need for this extra representation \cite{Mohapatra:1982tc} 
is due to the fact that  
$\phi_{\overline{126}}$ 
breaks $\U(1)_{R} \times \U(1)_{B-L} \times \U(1)_{\rm PQ} \to \U(1)'_{\rm PQ} \times \U(1)_Y$, 
where $\U(1)'_{\rm PQ}$ is a remnant PQ symmetry (a linear combination of the 
original PQ and the broken gauge generators) that 
would be eventually broken 
at the electroweak scale, leading to a phenomenologically untenable 
Weinberg-Wilczek axion. 
Hence, the $\phi_{16}$ (together with $\phi_{\overline{126}}$) 
is needed to ensure a proper rank reduction down to the SM group. 
In order for the $\phi_{16}$ to participate to $\U(1)_{\rm PQ}$ breaking, 
it needs to couple in a non-trivial way 
to $\phi_{10}$ and/or $\phi_{\overline{126}}$ 
so that it can get 
charged under the PQ. 
The simplest, viable option is 
that $\phi_{16} \sim \overline{\3}$, 
thus allowing for the gauge invariant operator 
$\phi^2_{16} \phi^\star_{10}(\phi_{45})$
(the parenthesis meaning both the invariant 
with and without $\phi_{45}$), 
so that $\PQ(\phi_{16}) = -1$.\footnote{\label{foot:16align1}Refs.~\cite{Mohapatra:1982tc,Chang:1987hz} considered 
instead the case in which the following 
operators 
$\phi^2_{16} \phi_{10}(\phi_{45})$ and
$\phi^2_{16} \phi_{\overline{126}} (\phi_{45})$ 
are present, and hence 
$\PQ(\phi_{16}) = 1$.
As we are going to show in \sect{sec:physaxion} (cf.~footnote (\ref{foot:align})),  
the latter choice implies an alignment between 
$\vev{\phi_{16}}$ and $\vev{\phi_{\overline{126}}}$, ending up in a 
Weinberg-Wilczek axion.}
\end{itemize}
Summarizing, the Higgs sector includes $\phi_{10} \sim \overline{\textbf{6}}$, $\phi_{\overline{126}} \sim \overline{\textbf{6}}$, 
$\phi_{45} \sim \textbf{1}$ and $\phi_{16}  \sim \overline{\3}$. 
Then the scalar potential, 
$\V = \V_2 + \V_3 + \V_4$ (with the subscript denoting the dimensionality of the operators), 
features the following 
$\SO(10) \times \SU(3)_f$ invariant terms 
\begin{align}
\label{eq:V2}
\V_2 &= |\phi_{10}|^2 + |\phi_{\overline{126}}|^2 + \phi_{45}^2 + |\phi_{16}|^2 \, , \\  
\label{eq:V3}
\V_3 &= \phi^2_{16} \phi^{\star}_{10} 
+ \text{h.c.}
\, , \\
\label{eq:V4}
\V_4 &= 
\V^2_2\text{-terms} +
\phi^2_{10} \phi^{\star 2}_{\overline{126}} 
+ \phi_{10} \phi_{\overline{126}} \phi^{\star 2}_{\overline{126}} 
+ \phi^2_{16} \phi^{\star}_{10} \phi_{45} 
+ \text{h.c.}
\, , 
\end{align}
where $\V^2_2\text{-terms}$ stands for quartics obtained by ``squaring'' $\V_2$ 
(including all possible 
linearly independent invariants made by the same amount of fields).  
Note that the operator 
$\phi^2_{16} \phi^{\star}_{\overline{126}} (\phi_{45})$ is not 
allowed by SO(10) invariance, 
but due to the interplay with the $\phi_{10}$ there is a 
sufficient amount of scalar potential terms 
so that a single abelian global symmetry survives 
accidentally. 
This can be identified with the $\U(1)_{\rm PQ}$,  
with transformation properties 
\beq 
\label{eq:PQVSO10}
\phi_{16} \to e^{-i\alpha} \phi_{16}  \, , \quad
\phi_{45} \to \phi_{45} \, ,
\eeq
while those of $\phi_{10}$ and $\phi_{\overline{126}}$ are given in \eq{eq:PQLY}.    
The $\SU(3)_f$ symmetry forbids the following terms which are allowed by SO(10) invariance:
\beq 
\! \phi_{10}^2 \, , \
\phi_{10}^4 \, , \
\phi_{16}^4 \, , \
\phi_{\overline{126}}^4 \, , \
\phi_{10}^2 \phi_{45}^2 \, , \
\phi^2_{10} \phi^{2}_{\overline{126}} \, , \
\phi_{10} \phi^{3}_{\overline{126}} \, , \ 
\phi_{45}^2 \phi_{\overline{126}}^2 \, , \
\phi^2_{16} \phi_{10} (\phi_{45}) \, , \
\phi^2_{16} \phi_{\overline{126}} (\phi_{45}) \, . 
\eeq
The origin of the accidental $\U(1)_{\rm PQ}$ can be 
neatly understood in terms of the action 
of the centers 
of $\SO(10)$ and $\SU(3)_f$, which are respectively 
$\mathbb{Z}_4$ and $\mathbb{Z}_3$.\footnote{The center $Z(G)$ of a group $G$ 
is the set of elements that commute with every element of $G$. 
$Z(\SU(3)) = \mathbb{Z}_3$, which is generated by $e^{2\pi i/3} \mathbbm{1}_3$; 
$Z(\SO(10)) = \mathbb{Z}_4$, which is generated by $i\Gamma_0$, 
with $\Gamma_0$ denoting the ``chirality'' operator
of the SO(10) Clifford algebra (see e.g.~\cite{Lazarides:1982tw}). 
} 
The transformation properties of the model fields under the 
gauge and the $\U(1)_{\PQ}$ accidental symmetries are summarized in \Table{tab:SOirrep}.

\begin{table}[t]
$$\begin{array}{c|c|cc|cc|c}
\rowcolor[HTML]{C0C0C0} 
\hbox{Field} & \hbox{Lorentz} &  \SO(10) 
& 
\mathbb{Z}_4 
& \SU(3)_f  & 
\mathbb{Z}_3 
& \U(1)_{\rm PQ} \\ \hline
\psi_{16} & (1/2,0) & 16 & i & \3 & e^{i 2\pi/3} & 1 \\ 
\rowcolor{CGray} 
\psi_{1}^{1,\ldots,16} & (1/2,0) & 1 & 1 & \overline{\3} & e^{i 4\pi/3} & 0 \\
\hline
\phi_{10} & (0, 0) & 10 & -1 & \6 & e^{i 2\pi/3} & -2 \\ 
\phi_{16} & (0, 0) & 16 & i & \overline{\3} & e^{i 4\pi/3} & -1 \\ 
\phi_{\overline{126}} & (0, 0) & \overline{126} & -1 & \6 & e^{i 2\pi/3} & -2 \\ 
\phi_{45} & (0, 0) & 45 & 1 & \1 & 1 & 0 \\ 
\end{array}$$
\caption{Field content of the model and relative transformation properties under 
$\SO(10) \times \SU(3)_f$, its $\mathbb{Z}_4 \times \mathbb{Z}_3$ center
and the accidental $\U(1)_{\rm PQ}$. 
In light gray, exotic fermions
which 
ensure $\SU(3)^3_f$ anomaly cancellation 
(cf.~\sect{sec:SU3anomaly}). 
}
\label{tab:SOirrep}
\end{table}%

Note that for consistency we have also introduced 16 (SO(10)-singlet) exotic fermions 
in the $\overline{\3}$ of $\SU(3)_f$, 
to ensure $\SU(3)^3_f$ anomaly cancellation. Their spectrum will be discussed 
in \sect{sec:flavonsec}, together with $\SU(3)_f$ breaking.

\section{Peccei-Quinn quality}
\label{sec:PQquality}

After having obtained 
the $\U(1)_{\rm PQ}$ to arise accidentally in the renormalizable Lagrangian, 
one should worry about possible sources of PQ breaking in the UV, which are often 
parametrized via effective operators suppressed by a cut-off scale $\Lambda_{\rm UV}$. 
A simple estimate shows that $\U(1)_{\rm PQ}$ should be preserved 
by operators up to dimension $d \gtrsim 9$, 
assuming for instance $\Lambda_{\rm UV} \sim M_{\rm Pl}$ 
and an 
axion decay constant $f_a \sim 10^9$ GeV. 
This is obtained 
by requiring that the energy density from UV sources of PQ breaking is 
about 
$10^{-10}$ 
times smaller than the energy density 
of the QCD axion potential
\beq 
\label{eq:PQestimate}
\( \frac{f_a}{\Lambda_{\rm UV}} \)^{d-4} f_a^4 \lesssim 10^{-10} \Lambda^4_{\rm QCD} \, ,
\eeq
so that the induced axion VEV 
displacement from zero 
is $\vev{a} / f_a \lesssim 10^{-10}$, 
within the bound from the 
neutron electric dipole moment (nEDM). 

Interestingly, the gauging of $\SU(3)_f$ provides some protection 
also beyond the renormalizable level. Given the field content in 
\Table{tab:SOirrep}, we proceed to identify at which operator level the 
$\U(1)_{\rm PQ}$ gets broken in the scalar potential. 
The lowest-dimensional PQ-breaking operators 
are found to be: 
\begin{align}
&\phi_{10}^6 & &(d=6) \, , \\
&\phi_{\overline{126}}^6 & &(d=6) \, , \\
&\phi_{16}^6 \phi^3_{10} & &(d=9) \, , \\
&\phi_{16}^6 \phi^3_{\overline{126}} & &(d=9) \, , \\ 
&\phi_{16}^{12} & &(d=12) \, . 
\end{align}
This classification can be easily understood in terms of the action of the 
$\mathbb{Z}_4 \times \mathbb{Z}_3$ center as displayed in \Table{tab:SOirrep}: 
invariance under $\mathbb{Z}_3$ requires a number of fields 
that is a multiple of 3 and after that one has to 
compensate powers
in order to get a $\mathbb{Z}_4$ singlet. 
Note that the $d=9$ operators 
$\phi_{16}^6 \phi_{10}^{\star 3}$ and $\phi_{16}^6 \phi_{\overline{126}}^{\star 3}$
are also allowed by gauge invariance, but they preserve $\U(1)_{\rm PQ}$. 

Some comments are in order, regarding the impact of those operators on the 
PQ quality problem: 
\begin{itemize}
\item $\phi_{10}$ can only develop electroweak scale VEVs, 
hence the Planck-suppressed 
operators
$\phi^6_{10}$ and $\phi_{16}^6 \phi^3_{10}$ 
do not pose a problem for the PQ quality issue.
Then, we only need to worry about operators developing 
``large'' VEVs compared to the 
electroweak scale, i.e.~$\phi_{16}$ and $\phi_{\overline{126}}$. 
\item Remarkably, the operators $\phi_{\overline{126}}^6$ and $\phi_{16}^{12}$ 
do not yield a large contribution to the axion potential, 
since (once projected on the SM vacuum) extra electroweak 
VEV insertions are needed. 
To see this, recall the decompositions under 
$\SO(10) \to \SU(5) \times \U(1)_Z$ \cite{Slansky:1981yr}: 
$16 \to 1(-5) + \ldots $ and $\overline{126} \to 1(+10) + \ldots$, 
along the $\SU(5)$ singlet components. 
Clearly, $\phi_{\overline{126}}^6$ and $\phi_{16}^{12}$ have zero 
projection on $V_{\overline{126}}^6$ and $V_{16}^{12}$ 
(the latter denoting the SU(5)-singlet VEVs), 
since it is not possible to compensate the $\U(1)_Z$ charge.   

\item The leading contribution to the axion potential is then given by the operator 
$\phi_{16}^6 \phi^3_{\overline{126}}$, with a non-zero 
projection on the high-scale VEVs $V_{16}^{6} V_{\overline{126}}^3$. 
In order to assess the impact of that PQ-breaking operator on the axion potential, 
one has still to identify the physical axion field. This step will be done in 
detail \sect{sec:physaxion}, but we anticipate here 
the results for the assessment of the PQ quality problem. 
Using \eq{eq:decphi16bis} we find the following 
contribution to the axion potential: 
\beq 
\label{eq:PQbreakPQquality}
\V_{\rm PQ-break} = k \frac{\phi_{16}^6 \phi^3_{\overline{126}}}{M^5_{\rm Pl}} + \text{h.c.} 
\supset \frac{|k|}{2^{7/2}} \frac{V_{16}^6 V^3_{\overline{126}}}{M^5_{\rm Pl}} 
\cos\(  \frac{ a}{V/12} + \delta \)
\, ,  
\eeq
with $\delta = \text{Arg}\, k$ denoting a generic $\mathcal{O}(1)$ phase and 
$V$ a function of $V_{16,\overline{126}}$ defined in \eq{eq:defV} (for $c_1 = 1$). 
Considering the 
total axion potential $\V = \V_{\rm PQ-break} + \V_{\rm QCD}$, 
with $\V_{\rm QCD} = - \Lambda_{\rm QCD}^4 \cos (a /f_a)$
and $f_a$ given in terms of $V_{16,\overline{126}}$ via \eq{eq:axiondc}, 
one finds the induced QCD theta term (valid for $\vev{a} / f_a \ll 1$): 
\beq 
\label{eq:thetain} 
\theta_{\rm eff} \equiv \frac{\vev{a}}{f_a} \approx 
\frac{|k| V_{16}^6 V^3_{\overline{126}} \sin\delta}
{8 \sqrt{2} M^5_{\rm Pl} \Lambda_{\rm QCD}^4 } 
\, . 
\eeq
Requiring the neutron electric dipole moment bound, $|\theta_{\rm eff}| 
\lesssim 2.0 \times 10^{-10}$, 
implies 
$(V^3_{16} V^6_{\overline{126}})^{1/9} \lesssim 2.0 \times 10^{9}$ GeV 
(for e.g.~$|k| = 1$ and $\sin\delta = 1$). A more useful estimate in terms 
of the axion decay constant will be given in \sect{sec:axionexp}.

\item Last, but not least, assuming no new physics above the GUT scale 
apart for gravity, 
it is unclear whether quantum gravity
would generate power-like PQ-breaking operators, 
such as those in \eq{eq:PQbreakPQquality}. 
Nonetheless, it is believed that gravity does violate global symmetries, 
based on semi-classical arguments related to black holes and Hawking radiation. 
In scenarios in which Einstein gravity is minimally
coupled to the axion field, non-conservation of the PQ global charge arises 
from non-perturbative effects described by Euclidean wormholes.
Those are calculable in the semi-classical limit \cite{Abbott:1989jw,Coleman:1989zu,Kallosh:1995hi,Alonso:2017avz,Hebecker:2018ofv} and give a correction to the axion potential of the order of
$M_{\rm Pl}^4 e^{-S_{\rm wh}}$, 
where $S_{\rm wh} \sim M_{\rm Pl} / f_a$ is the wormhole action.  
Being exponentially suppressed, such contribution 
poses a problem for the PQ solution only for $f_a \gtrsim 10^{17}$ GeV, 
which is however much above the intermediate-scale values of $f_a$
that will be considered here. 
\end{itemize} 
All in all, we conclude that not only the gauging of $\SU(3)_f$ leads to an accidental 
$\U(1)_{\rm PQ}$, but it also structurally protects it against 
PQ breaking 
effective operators. 
To solve the PQ quality issue in its standard formulation in terms of 
Planck-suppressed, power-like 
effective operators one needs PQ breaking 
VEVs smaller than about $10^{9}$ GeV. 

\section{Axion's guts (physical axion field)}
\label{sec:physaxion}

A crucial quantity that qualifies the U(1) accidental symmetry as a 
PQ symmetry 
is the QCD 
anomaly of the classically conserved PQ current, $J^{\rm PQ}_{\mu}$. 
At the SO(10) level,  
that is 
(employing the compact notation 
$G \tilde G \equiv G^a_{\mu\nu} \tilde G^{a\, \mu\nu}$, etc.)
\begin{align}
\label{eq:dJPQAnomalSO10}
\partial^\mu J^{\rm PQ}_{\mu} &=
\frac{\alpha_{\SO(10)} N_{\SO(10)}}{4\pi} F_{\SO(10)} \tilde F_{\SO(10)} \nonumber \\
&\supset  
\frac{\alpha_s N_{\SO(10)}}{4\pi} 
G \tilde G
+\frac{\alpha_W N_{\SO(10)}}{4\pi} 
W \tilde W
+\frac{5}{3} \frac{\alpha_Y N_{\SO(10)}}{4\pi} 
B \tilde B \nonumber \\
&\supset  
\frac{\alpha_s N_{\SO(10)}}{4\pi} 
G \tilde G
+\frac{8}{3} \frac{\alpha N_{\SO(10)}}{4\pi} 
F \tilde F
\equiv 
\frac{\alpha_s N}{4\pi} 
G \tilde G
+ \frac{\alpha E}{4\pi} 
F \tilde F
\, , 
\end{align}
where the matching with the SM gauge bosons 
is understood at the GUT scale. 
The matching with QED is obtained by projecting on the photon component 
($W^3 \to \sin\theta_W A$ and $B \to \cos\theta_W A$) 
and using $\alpha = \alpha_W \sin^2\theta_W = \alpha_Y \cos^2\theta_W$.
In the last step we made contact with the standard notation 
for the QCD ($N$) and QED ($E$) anomaly coefficients, which read respectively
$N = N_{\SO(10)}$ and $E = 8/3 \, N_{\SO(10)}$,  
in terms of the SO(10) anomaly factor 
\beq 
\label{eq:SO10anomalyf}
N_{\SO(10)} = n_g T(\psi_{16}) \PQ(\psi_{16}) = 6 \, ,
\eeq
with
$n_g = 3$ (number of generations),  
$T(\psi_{16}) = 2$ (Dynkin index of the spinorial of SO(10)) 
and $\PQ(\psi_{16}) = 1$.  
Upon the anomalous PQ rotation $\psi_{16} \to e^{i\alpha} \psi_{16}$, the QCD 
theta term in $\mathscr{L}_{\rm QCD} \supset \frac{\alpha_s \theta}{8 \pi} G \tilde G$ 
is shifted as $\theta \to \theta + 2N \alpha $. Since different values 
of $\theta$ correspond to different physics modulo $2\pi$, 
the $\U(1)_{\PQ}$ symmetry is explicitly broken down to a 
$\mathbb{Z}_{2N}$, corresponding to $\alpha = 2\pi n / (2N)$ 
with $n=0,1,\ldots,2N-1$. 

The anomaly does not depend on the mass of the fermions running in the 
triangle loop, and hence it 
must be preserved through the various stages of symmetry breaking. 
This feature will be useful for identifying the physical axion field 
and its low-energy couplings. 
In the following, 
we provide the anatomy of the axion field in SO(10)
at three different levels of increasing complexity: 
$i)$ a single, dominant SM-singlet and PQ-breaking VEV, 
$ii)$ two SM-singlet and PQ-breaking VEVs of comparable size 
and $iii)$ the full-fledged case including also electroweak VEVs, 
a necessary step to compute low-energy axion couplings to SM matter fields. 
The last, self-contained derivation is deferred to \app{sec:loweaxion}.

\subsection{A dominant SM-singlet and PQ-breaking VEV}
\label{sec:onedomVEV}

Let us first consider a phenomenologically relevant limit where physics should be clearer, 
and focus on the breaking pattern 
\begin{align}
\label{eq:SO10breakpatt}
\SO(10) \times \U(1)_{\rm PQ} & \xrightarrow[]{\vev{\phi_{45}}_{B-L}} 
\SU(3)_c \times \SU(2)_L \times \SU(2)_R \times \U(1)_{B-L} \times \U(1)_{\rm PQ} \nonumber \\
& \xrightarrow[]{V_{\overline{126}}}  
\SU(3)_c \times \SU(2)_L \times \U(1)_{Y} \times \U(1)'_{\rm PQ} 
\nonumber \\
& \xrightarrow[]{V_{16}}  
\SU(3)_c \times \SU(2)_L \times \U(1)_{Y} 
\, , 
\end{align}
where we assumed the hierarchy of VEVs: 
$\vev{\phi_{45}}_{B-L} \gg V_{\overline{126}} \gg V_{16}$, 
as motivated by gauge coupling unification \cite{Chang:1984qr,Deshpande:1992em,Bertolini:2009qj}.  
Here, $\vev{\phi_{45}}_{B-L}$ denotes a specific orientation of the adjoint VEV, 
which can be achieved via a renormalizable 
scalar potential after including one-loop corrections \cite{Bertolini:2009es,Bertolini:2012im,Bertolini:2013vta,Graf:2016znk}; 
while the other two intermediate-scale VEVs are defined in terms of the 
following polar field decompositions (along the SM-singlet components)
\beq
\label{eq:decphi16}
\phi_{16} = \frac{1}{\sqrt{2}} ( V_{16} + \ldots ) e^{i \frac{a_{16}}{V_{16}}} \, , \qquad 
\phi_{\overline{126}} = \frac{1}{\sqrt{2}} ( V_{\overline{126}} + \ldots ) e^{i \frac{a_{\overline{126}}}{V_{\overline{126}}}} \, , 
\eeq
with the projections of the angular modes $a_{16}$ and $a_{\overline{126}}$ 
on the axion field yet to be identified. 
After the second breaking stage, 
a remnant $\U(1)'_{\rm PQ}$ 
global symmetry is left invariant by $\vev{\phi_{\overline{126}}}$ 
(which cannot reduce the rank by more than one unit). 
The latter symmetry can be expressed 
as a linear combination of the original $\U(1)_{\rm PQ}$ and the broken 
Cartan gauge generators as\footnote{Since at this point we only consider 
VEVs with zero hypercharge, $Y = T^3_R + (B-L)/2$, 
it is not necessary to include the $T^3_R$ generator.} 
\beq 
\label{eq:defPQp}
\text{PQ}' = c_1 \, \text{PQ} + c_2 (B-L) \, .  
\eeq
Given $\{\text{PQ}, B-L\} (\vev{\phi_{\overline{126}}}) = 
\{-2, -2\}$, the defining property, $\text{PQ}' (\vev{\phi_{\overline{126}}}) = 0$, 
implies $c_2 = - c_1$. 
The coefficient $c_1$ can be fixed by matching the anomaly  
in \eq{eq:SO10anomalyf} with the one computed in terms of $\PQ'$ charges, 
that is 
\beq 
\label{eq:NPQpchaarges}
N = n_g T(3) (2 \, \PQ' (q) + 2 \, \PQ' (q^c)) = 6 c_1 \, ,
\eeq
where we used $n_g = 3$, $T(3) = 1/2$ (the Dynkin index of the fundamental of $\SU(3)_c$) and the $\PQ'$ charges $\PQ'(q) = 2c_1/3$ 
(for the $\SU(2)_L$ quark doublet, with $B-L = 1/3$)
and $\PQ'(q^c) = 4c_1/3$ 
(for the $\SU(2)_R$ quark doublet, with $B-L = -1/3$). The matching with the 
SO(10) anomaly in \eq{eq:SO10anomalyf} 
requires then $c_1 = 1$, and hence 
$\PQ' = \PQ - (B-L)$. 
Given $\{\text{PQ}, B-L\} (\vev{\phi_{16}})  
= \{-1, 1\}$, it is readily verified that the action of $\text{PQ}'$ on $\vev{\phi_{16}}$ 
is non-trivial ($\text{PQ}' (\vev{\phi_{16}}) = -2$) 
and hence it allows us to consistently break $\U(1)'_{\rm PQ}$ 
at the scale $V_{16}$.\footnote{\label{foot:align}Choosing instead 
the PQ charge of $\phi_{16}$ equal to 1, leads to $\text{PQ}' \vev{\phi_{16}} = 0$.  
Namely $\vev{\phi_{16}}$ and $\vev{\phi_{\overline{126}}}$ are aligned, 
thus not providing a proper breaking pattern 
for $\U(1)'_{\rm PQ}$, 
that gets eventually broken at the electroweak scale.} 
Neglecting $(V_{16} / V_{\overline{126}})^2$ corrections as well 
as subleading electroweak VEVs,
the axion field is identified with the angular component of the 
$\phi_{16}$ field, $a = a_{16}$. 

In the other relevant limit, $V_{16} \gg V_{\overline{126}}$, 
one obtains that another linear combination 
$\U(1)''_{\text{PQ}}$, defined as
$\PQ'' = \PQ + (B-L)$, 
is left invariant by $\vev{\phi_{16}}$, 
which is further broken by $\vev{\phi_{\overline{126}}}$ 
($\PQ'' (\vev{\phi_{\overline{126}}}) = - 4$), 
so that the axion field is $a = a_{\overline{126}}$.

\subsection{Two SM-singlet and PQ-breaking VEVs}
\label{sec:twoVEVs}

The more general case with $V_{16} \sim V_{\overline{126}}$ 
must interpolate between the two limits above and it 
can be obtained as follows. Consider the classically conserved currents 
\begin{align}
\label{eq:JPQ}
J_\mu^{\rm PQ} &= 
q_{16} V_{16} \partial_\mu a_{16} + 
q_{\overline{126}} V_{\overline{126}} \partial_\mu a_{\overline{126}} \, , \\
\label{eq:JBmL}
J_\mu^{B-L} &= 
(B-L)_{16}
V_{16} \partial_\mu a_{16} + 
(B-L)_{\overline{126}}
V_{\overline{126}} \partial_\mu a_{\overline{126}} \, , 
\end{align}
with gauge charges $(B-L)_{16} = 1$ and $(B-L)_{\overline{126}} = -2$, 
and the \emph{physical} PQ charges $q_{16,\overline{126}}$. 
The latter are linear combination of the PQ charges in \Table{tab:SOirrep} 
and broken gauge generators, in general 
\beq 
\label{eq:generalUVIRPQ}
q = c_1 \PQ + c_2 (B-L) + c_3 Y \, , 
\eeq
(with $c_3 = 0$ as long as we consider only the SM-singlet VEVs 
$V_{16,\overline{126}}$). 
The $q$ charges can be fixed by requiring that the PQ 
and $B-L$ currents are orthogonal:  
\beq 
\label{eq:orthogPQBmL}
q_{16} V^2_{16} 
- 2 q_{\overline{126}} V^2_{\overline{126}}  = 0 \, , 
\eeq
which ensures no kinetic mixings between 
the axion field and the $B-L$ massive gauge boson,
thus providing a canonical axion field. The latter is defined as 
\beq 
\label{eq:axiondef}
a = \frac{1}{V} (q_{16} V_{16} a_{16} + 
q_{\overline{126}} V_{\overline{126}} a_{\overline{126}}) \, ,
\eeq
with 
\beq 
\label{eq:Vnorm}
V^2 = (q_{16})^2 V^2_{16} 
+ (q_{\overline{126}})^2 V^2_{\overline{126}} \, ,
\eeq
so that $J_\mu^{\rm PQ} = V \partial_\mu a$ and, 
compatibly with the Goldstone theorem, 
$\langle 0 | J_\mu^{\rm PQ} | a \rangle = i V p_\mu$.  
Inverting the orthogonal transformation in \eq{eq:axiondef}, 
one readily 
obtains the 
projection of the angular modes on the axion field: 
\beq 
\label{eq:aproj}
a_{16} \to q_{16} V_{16} \frac{a}{V} \, , \qquad 
a_{\overline{126}} \to q_{\overline{126}} V_{\overline{126}} \frac{a}{V} \, , 
\eeq 
whose weight factors are extracted from \eq{eq:orthogPQBmL} 
and (\ref{eq:Vnorm}), and can be conveniently written as 
\beq 
\label{eq:weightf}
q_{16} V_{16} = V \cos\omega \, , \qquad 
q_{\overline{126}} V_{\overline{126}} = V \sin\omega \, , 
\eeq
in terms of the vacuum parameter 
\beq 
\label{eq:defgamma}
\tan\omega = \frac{V_{16}}{2V_{\overline{126}}} \, . 
\eeq 
So we can finally read the axion composition 
of the complex fields in 
\eq{eq:decphi16}, which is 
\beq
\label{eq:decphi16bis}
\phi_{16} =
\frac{1}{\sqrt{2}} ( V_{16} + \ldots )
e^{i \cos\omega \frac{a}{V_{16}}} \, , \qquad 
\phi_{\overline{126}} =
\frac{1}{\sqrt{2}} ( V_{\overline{126}} + \ldots )
e^{i \sin\omega \frac{a}{V_{\overline{126}}}} \, ,   
\eeq
For instance, taking the limit $V_{\overline{126}} \gg V_{16}$ ($\omega \to 0$) 
one reproduces the previous result that the axion dominantly 
correspond to the angular mode of $\phi_{16}$.

The orthogonality condition in \eq{eq:orthogPQBmL}, 
together with the 
matching of the UV and IR charges of $\phi_{16}$ and 
$\phi_{\overline{126}}$ in \eq{eq:generalUVIRPQ}, allows us to fix 
\beq 
c_2 = - c_1 \cos2\omega \, , \qquad 
q_{16} = - 2 c_1 \cos^2\omega  \, , \qquad
q_{\overline{126}} = - 4 c_1 \sin^2\omega  \, ,
\eeq
and hence, using \eq{eq:Vnorm} (and \eq{eq:weightf} regarding the absolute sign)
\beq 
\label{eq:defV}
V = - \frac{4 c_1 V_{\overline{126}} V_{16}}{\sqrt{V^2_{16} + 4 V^2_{\overline{126}}}} \, .
\eeq
Again, the value of $c_1$ can be determined by matching the anomaly 
between the UV and the broken theory, that is 
\beq 
\label{eq:NPQpchaarges2}
N = n_g T(3) (2 \, q(q) + 2 \,  q(q^c)) = 6 c_1 \, ,
\eeq
where $q(q) = 2c_1/3 (\cos^2\omega + 2\sin^2\omega)$ 
and 
$q(q^c) = 2c_1/3 (2\cos^2\omega + \sin^2\omega)$.  
Hence, $c_1 = 1$ in order to match the UV anomaly. 
The correct physical limits 
$V = -2 V_{16}$ (for $V_{\overline{126}} \to \infty$) and 
$V = -4 V_{\overline{126}}$ (for $V_{16} \to \infty$) are then 
recovered.\footnote{To see that, 
take for instance the $V_{\overline{126}} \to \infty$ limit, 
corresponding to $a \to a_{16}$. 
Then, given 
that under a PQ transformation $a_{16,\overline{126}} \to a_{16,\overline{126}} 
+ \alpha q_{16,\overline{126}} V_{16,\overline{126}}$ 
the axion field transforms as $a \to a + \alpha V$ (cf.~\eq{eq:axiondef}),  
we can make the identification  
$V = q_{16} V_{16} \to -2 V_{16}$.} 
Finally, the axion decay constant is 
\beq 
\label{eq:axiondc}
f_a = \frac{V}{2N} = -
\frac{V_{\overline{126}} V_{16}}{3\sqrt{V^2_{16} + 4 V^2_{\overline{126}}}} 
\, . 
\eeq
In the following, to ease the notation when discussing phenomenological bounds, 
we will refer to $f_a$ meaning its absolute value. 

The most general case including also electroweak VEVs 
that participate to $\U(1)_{\PQ}$ breaking is discussed in \app{sec:loweaxion}, 
together with the derivation of low-energy axion couplings.

\subsection{Axion domain wall problem?}
\label{sec:axionDW}

The discrete $\mathbb{Z}_{2N}$ symmetry left invariant by the QCD 
anomaly (cf.~discussion below \eq{eq:SO10anomalyf}), implies 
that the axion potential has $2N=12$ degenerate minima. 
At the 
QCD phase transition, 
this leads to the formation of domain-wall-like structures 
at the boundaries between regions of different vacua, which
quickly 
dominate the energy density 
of the Universe.  
However, in the presence of extra 
global/local symmetries it could happen 
that some of those minima are connected. 
In particular, if the $\mathbb{Z}_{2N}$ 
discrete symmetry 
can be fully embedded in the center of a non-abelian symmetry group, then  
the axion domain walls will quickly disappear through 
the emission of Goldstone bosons, thus providing an elegant 
solution to the axion domain wall 
problem \cite{Lazarides:1982tw}. 
Solutions of this type have been 
often considered in the context of 
SO(10) \cite{Lazarides:1982tw,Barr:1982bb,Lazarides:2020frf}, 
whose $\mathbb{Z}_{4}$ center provides a starting point for embedding  
the $\mathbb{Z}_{2N}$ discrete symmetry.   
This is possible in models where $2N=4$ 
or if extra global/local symmetries beyond SO(10) are invoked 
(which is the case considered here).

While such a solution was claimed to be at play also in the 
$\SO(10) \times \SU(3)^{\rm global}_f$ model of Ref.~\cite{Chang:1987hz}, 
sharing several similarities with the present work, 
it turns out that this 
conclusion is flawed by the fact that in the model of \cite{Chang:1987hz} 
the PQ symmetries of 
$\phi_{16}$ and $\phi_{\overline{126}}$ were aligned, 
eventually leading to a phenomenologically untenable 
Weinberg-Wilczek axion (cf.~the discussion 
in footnote (\ref{foot:16align1})). 
After assigning the $\phi_{16}$ to a proper $\SU(3)_f$ representation 
(cf.~\Table{tab:SOirrep}), 
so that its PQ charge is not aligned to the one of $\phi_{\overline{126}}$, 
it turns out that the remnant $\mathbb{Z}_{2N}$ 
symmetry cannot be embedded in the center 
of $\SO(10) \times \SU(3)_f$ and hence the 
axion domain wall problem persists.  
To show this, let us define the action of the centers 
as $r_P = \exp{(i2\pi/P)}$. 
Denoting as $r_{12}$ the action of a PQ transformation with discrete parameter 
$\alpha = 2\pi / 12$, $r_4$ and $r_3$ 
the actions of the $\SO(10)$ and $\SU(3)_f$ centers 
(as displayed in \Table{tab:SOirrep}),  
one would like to have that $r_{12} = r^{-1}_4 r_3$ on all the model fields, 
so that the two sets of symmetries can be identified.
It is easy to check that while such identification 
works for the fields $\psi_{16}$ ($r_{12} = r^{-1}_4 r_3 = \exp{(i\pi/6)}$), 
$\phi_{10}$, $\phi_{\overline{126}}$ ($r_{12} = r^{-1}_4 r_3 = \exp{(-i\pi/3)}$)
and $\phi_{45}$ ($r_{12} = r^{-1}_4 r_3 = 1$), 
it fails for 
$\phi_{16}$ ($r_{12} = \exp{(-i\pi/6)} \neq r^{-1}_4 r_3 = \exp{(i5\pi/6)}$) 
and 
$\psi_{1}^{1,\ldots,16}$ ($r_{12} = 1 \neq r^{-1}_4 r_3 = \exp{(i2\pi/3)}$).

We hence conclude that the accidental SO(10) axion 
has a genuine domain wall problem, with domain wall 
number $N_{\rm DW} \equiv 2N = 12$.  
A straightforward solution is given by the pre-inflationary 
PQ breaking scenario (in which domain walls are inflated away). 
Instead, in the post-inflationary PQ breaking scenario 
the Planck-suppressed 
sources of explicit PQ breaking 
discussed in \sect{sec:PQquality} could potentially 
lift the vacuum degeneracy and 
lead to a fast decay 
of the domain walls \cite{Sikivie:1982qv,Kawasaki:2014sqa}, 
compatibly with the PQ solution of the strong CP problem.   
This possibility will be explored in more detail in \sect{sec:axionexp}.

\section{$\SU(3)_f$ sector}
\label{sec:flavonsec}

We next discuss two issues related to the $\SU(3)_f$ sector, 
namely the cancellation gauge anomalies and the spontaneous symmetry breaking 
of $\SU(3)_f$. 

\subsection{$\SU(3)^3_f$ anomaly}
\label{sec:SU3anomaly}

The only relevant gauge anomaly to be cancelled is the $\SU(3)^3_f$ one 
(while SO(10) is anomaly free).  
The simplest way to cancel the latter is to introduce 
16 exotic fermions, $\psi_1^{\alpha}$ ($\alpha=1,\ldots,16$)
which are SO(10) singlets and transform in the $\overline{\textbf{3}}$ of 
$\SU(3)_f$.\footnote{Other options to cancel the $\SU(3)^3_f$ anomaly 
include the following copies of SO(10)-singlet representations: 
$2 \times \overline{\mathbf 6} + 2 \times \overline{\mathbf 3}$, 
$3 \times \overline{\mathbf 6} + 5 \times \mathbf 3$, etc.} 
Due to their quantum numbers they have no renormalizable interactions 
with the other fields of \Table{tab:SOirrep}, so that they also acquire no PQ charge. 
Their spectrum is eventually controlled by $\SU(3)_f$ breaking, as discussed 
in the following. Let us also note that there is a remnant 
$\U(1)_{\rm PQ} \times \SU(3)_f^2$ anomaly 
(since only $\psi_{16}$ are charged under the PQ).  
So, in order for the axion to relax to zero the $\theta$ term of QCD 
(and not the one of $\SU(3)_f$), 
$\SU(3)_f$ should be spontaneously 
broken (effectively suppressing the contribution to the axion potential due to the 
Higgsing of $\SU(3)_f$ instantons). This is anyway compatible with the requirement that 
the flavour symmetry must be completely broken in order to give mass to SM fermions.  

\subsection{$\SU(3)_f$ breaking}
\label{sec:SU3breaking}

We assume that $\SU(3)_f$ breaks completely, in two steps 
\beq 
\SU(3)_f \xrightarrow[]{M_1} \SU(2)_f \xrightarrow[]{M_2} 
\mathbf{1}
\, , 
\eeq
where $\mathbf{1}$ denotes the trivial group ($\SU(2)_f$ completely broken). 
As suggested by SM charged fermion masses, 
we also assume 
a hierarchy of scales $M_1 \gg M_2$. 
This could be achieved 
via two (misaligned) scalars transforming 
as $\phi^{1,2} \sim \overline{\3}$ or $\mathbf{6}$ of $\SU(3)_f$. 
At the same time, Yukawa
operators  
$\psi_1^{\alpha} \psi_1^{\alpha} \phi^{1,2}$, 
$\psi_1^{\alpha} \psi_1^{\alpha} (\phi^{1,2})^2 / M_{\rm Pl}$,  
etc.~(in the diagonal basis), 
would give a mass of order $M_{1,2}$, $M^2_{1,2} / M_{\rm Pl}$, etc.~to the exotic fermions 
after complete $\SU(3)_f$ breaking. 
However, if $\phi^{1,2}$ were to transform under an SO(10) 
representation with trivial center, i.e.~1, 45, 54 or 210, 
the following operators would be also allowed: 
$\phi_{10}^2(\phi^{1,2})^{\star}$, 
$\phi_{10}^2(\phi^{1,2})^{\star 2}$ 
and $(\phi^{1,2})^3$, 
whose simultaneous presence would break explicitly the $\U(1)_{\PQ}$ 
at the renormalizable level. 
This can be avoided if we assigned instead $\phi^{1,2}$ e.g.~to a 16 or $\overline{126}$. 

A remarkably economical possibility 
(that is also unavoidable in absence of extra sources of $\SU(3)_f$ breaking at the GUT 
scale)
is actually that of using 
the representations $\phi_{16}$ and $\phi_{\overline{126}}$ in \Table{tab:SOirrep} 
to break simultaneously $\SO(10) \times \U(1)_{\PQ}$
\emph{and} $\SU(3)_f$. There are no 
obvious obstructions to this program, 
which ties together the ``horizontal'' and ``vertical'' breaking,
since the representations involved have the group-theoretical 
power to properly reduce the symmetry,\footnote{For instance, 
it should be clear that the VEV orientation 
$\vev{\phi_{16}}^{\alpha=16}_{a=3}$
breaks 
$\SO(10) \times \SU(3)_f \to \SU(5) \times \SU(2)_f$, etc.} 
while their high-scale 
VEVs would allow us to decouple the associated flavour breaking dynamics 
at safely large scales. 

On the other hand, within such approach the 16 
exotic fermions $\psi^{\alpha}_1$ remain massless at the renormalizable level.
The following 
effective operators, involving a pair of exotic fermions  
are allowed by $\SO(10) \times \SU(3)_f$ invariance 
(omitting the $\alpha=1,\ldots,16$ index)
\beq
\label{eq:mirroreffop}
\frac{1}{M_{\rm Pl}} \psi_1 \psi_1 \phi^2_{10} \, , \qquad 
\frac{1}{M^2_{\rm Pl}} \psi_1 \psi_1 \phi^2_{16} \phi_{\overline{126}} \, , \qquad 
\frac{1}{M^3_{\rm Pl}} \psi_1 \psi_1 \phi^4_{16} \, , \qquad 
\frac{1}{M^3_{\rm Pl}} \psi_1 \psi_1 (\phi^\star_{\overline{126}})^4 \, . 
\eeq
Only the second operator gives a sizeable contribution 
to the mass of the exotic fermions, that are lifted up to  
$(V^2_{16} V_{\overline{126}}) / M_{\rm Pl}^2 \lesssim 10$ TeV 
(for $V_{16,\overline{126}} \lesssim 10^{14}$ GeV), 
while the other operators have only projections on electroweak VEVs, 
thus providing a tiny contribution to the mass of the exotic fermions.\footnote{One 
might worry 
that the operators in \eq{eq:mirroreffop} break the PQ symmetry and hence do contribute 
to the axion potential upon closing fermion loops. 
Note, however, that it is possible to make at least one of those operators formally PQ 
invariant, by a proper assignment of the PQ charge of the exotic fermions, 
respectively $\PQ(\psi_1) = \{-2,2,2,-4\}$, 
so that the simultaneous presence of at least two of them 
is required in order to genuinely break the PQ symmetry. 
The largest contribution comes by combining the second operator, 
involving a SM-singlet VEV, with another operator that 
necessarily involves an electroweak scale VEV. 
The latter suppression makes the contribution to the 
axion potential safely negligible.} 
For lower values of $V_{16,\overline{126}}$ 
the exotic fermions are lighter
and they might eventually 
contribute to dark radiation if once in thermal contact with the SM 
via $\SU(3)_f$ interactions. 

Of course, it would be interesting to study whether 
the above flavour dynamics could be predictive for reproducing 
SM fermion masses and mixings. 
Here, we stress that the present approach to flavour 
differs from more standard ones (such as e.g.~the one of Ref.~\cite{Berezhiani:2005tp}), 
in which the Yukawas transform under $\SU(3)_f$, 
being themselves flavon fields interacting with SM fields via effective operators.
In our case instead it is crucial, in order to obtain an automatic $\U(1)_{\PQ}$, 
that the flavour dynamics acts at the renormalizable level and that 
a single Higgs representation breaks both $\SO(10)$ and $\SU(3)_f$,  
which is closer in spirit to the approach of Ref.~\cite{Joyce:1993vq}.  
A quantitative analyses of fermion masses and mixings 
following this path is left for future studies.

\section{Axion phenomenology}
\label{sec:axionexp}

\begin{figure}[t]
\centering
\includegraphics[width=16cm]{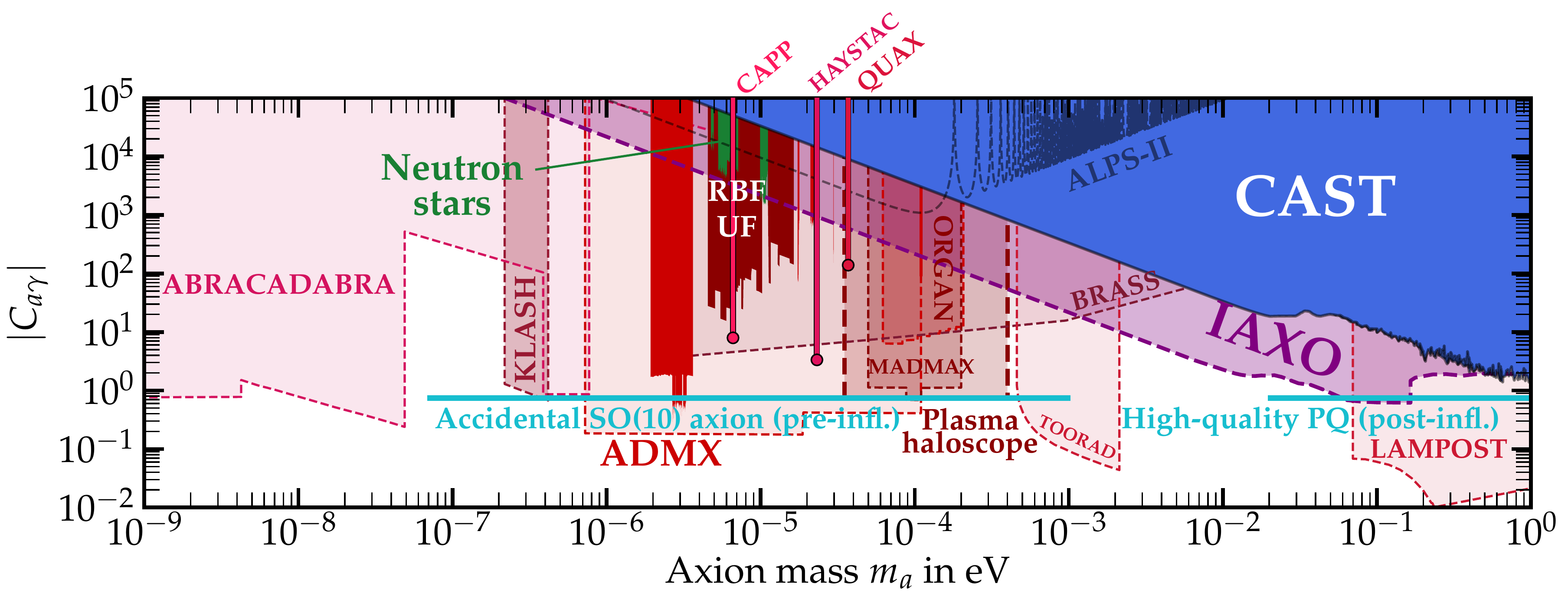}
\caption{Axion-photon coupling (\eq{eq:axionph} with $E/N = 8/3$)
and sensitivity of present (full lines) and future (dashed lines) axion experiments. 
The accidental SO(10) axion mass window 
(in the pre-inflationary PQ breaking scenario)
corresponds to $m_a \in [7 \times 10^{-8}, 10^{-3}]$ eV. 
A high-quality PQ symmetry is obtained for $m_a \gtrsim 0.02$ eV
(in the post-inflationary PQ breaking scenario). 
Axion limits from \cite{ciaran_o_hare_2020_3932430}.}
\label{fig:axionphoton}       
\end{figure}

In this Section we describe the phenomenological profile 
of the accidental SO(10) axion. Axion couplings to photons and SM matter 
fields have been computed in 
\app{sec:loweaxion} (respectively \eq{eq:axionph} and \eqs{eq:axioncu}{eq:axioncde}). 
For the present scenario, 
the axion coupling to photons 
represents the main experimental probe (as shown in \fig{fig:axionphoton}). 
The mass range of the accidental SO(10) axion 
is constrained by various considerations, 
depending on whether the PQ symmetry is broken before or after inflation. 

\subsection{Pre-inflationary PQ breaking}

If the PQ symmetry is broken before (and during) inflation and not restored afterwards, 
the dark matter (DM) relic density via the vacuum misalignment mechanism is  
given by \cite{Borsanyi:2016ksw,Tanabashi:2018oca}
\beq 
\Omega_a h^2 = 0.12 \( \frac{f_a}{9 \times 10^{11} \ \text{GeV}} \)^{1.165} \theta_{\rm in}^2 \, ,
\eeq
which is valid for $\theta_{\rm in} \lesssim 1$. 
In the non-linear regime, $\theta_{\rm in} \to \pi$, 
DM axions are allowed to have masses as high as $m_a \lesssim 1$ meV, 
but no higher because quantum fluctuations during inflation 
would imply too large iso-curvature fluctuations \cite{Wantz:2009it}. 

On the other hand, a lower bound on $m_a$ follows as well 
within the model, since $f_a$ is bounded from above by the seesaw scale. 
Rewriting \eq{eq:axiondc} as 
$|f_a| = |\sin 2\omega| V_{B-L}  / 12$,
with $V_{B-L} \equiv \sqrt{V^2_{16} + 4 V^2_{\overline{126}}}$, 
we find the upper limit $f_a \leq V_{B-L} / 12$. 
Note that the mass of the $B-L$ gauge boson 
will be proportional to $V_{B-L}$ 
(because of the relative $B-L$ charges of the two VEVs).  
Hence, $V_{B-L}$ 
assumes the
meaning of $B-L$ breaking scale, 
subject to constraints both
from gauge coupling unification
and neutrino masses.   
In particular, right-handed neutrino 
masses are controlled at tree-level by $V_{\overline{126}}$, 
via $M_{\nu_R} = y_{\overline{126}} V_{\overline{126}}/\sqrt{2}$, 
and at two loops \cite{Witten:1979nr,Bajc:2004hr}
by $V_{16}$. 
Very conservatively, we take $V_{B-L} \lesssim 10^{15}$ GeV, 
corresponding to $m_a \gtrsim 7 \times 10^{-8}$ eV.

\fig{fig:axionphoton} shows the predictions of the model 
in the axion-photon vs.~axion-mass plane. 
Barring a hole in sensitivity around $m_a \sim 10^{-7}$ eV, 
the parameter space of the accidental SO(10) axion 
in the pre-inflationary PQ breaking scenario 
will be mostly explored 
in the coming decades, under the crucial assumption that the 
axion comprises the whole DM (otherwise the 
sensitivity of axion DM experiments   
on the axion-photon coupling is diluted as 
$(\Omega_a / \Omega_{\rm DM})^{1/2}$).

It should be noted that although the allowed axion mass window 
spans over several orders of magnitude, 
further refinements of the present analysis will likely narrow down 
the axion mass range. This is due to extra constraints on the $B-L$ and PQ 
breaking scales originating from the requirement of reproducing SM fermion 
masses and mixings as well as from gauge coupling unification. 
However, we refrain from applying standard unification constraints to the present model, 
since those are affected by the replication of the SO(10) Higgs fields under $\SU(3)_f$, 
which could drastically change the running \cite{Banerjee:2020ule}, 
depending on the pattern of $\SU(3)_f$ breaking.

\subsection{Post-inflationary PQ breaking}

\begin{figure}[t]
\centering
\includegraphics[width=8cm]{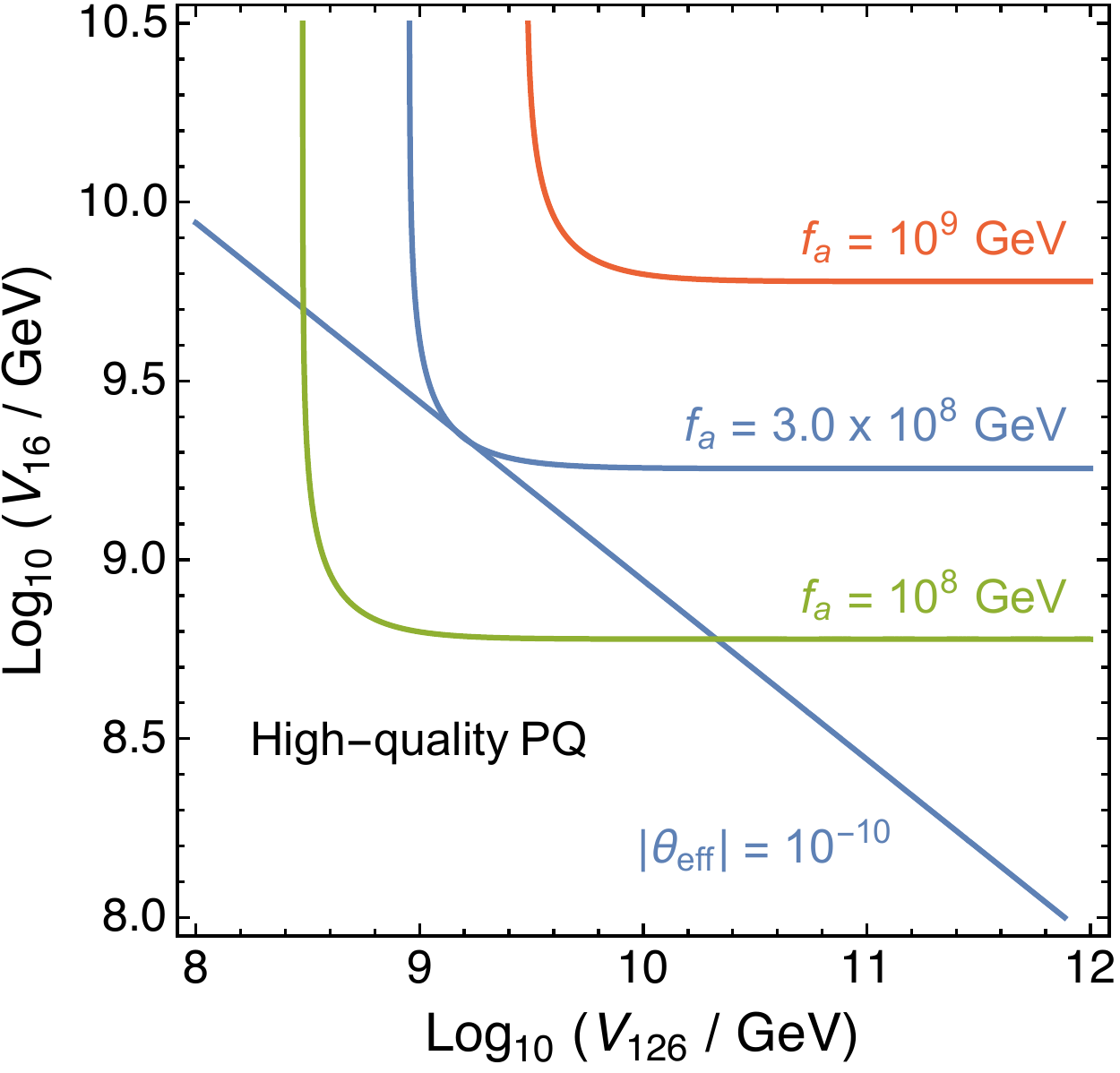}
\caption{Parameter space of the accidental SO(10) axion 
leading to a high-quality $\U(1)_{\PQ}$. For the estimate of 
the induced QCD theta term 
we used \eq{eq:thetain} with $|k| = 1$ and $\sin\delta = 1$.}
\label{fig:HQPQ}       
\end{figure}

If the PQ is broken after inflation or restored afterwards, 
one has also topological defects (axion strings and domain walls) 
that contribute to the axion DM relic density, on top of the 
usual misalignment mechanism. 
Such a scenario is theoretically motivated by the possibility of 
addressing the PQ quality problem in the present framework. 
In fact, using the estimate in \eq{eq:thetain} for the induced 
QCD theta term and the expression of $f_a$ in terms of $V_{16,\overline{126}}$ 
(cf.~\eq{eq:axiondc}), we find that in order to satisfy the nEDM 
bound, $|\theta_{\rm eff}| \lesssim 10^{-10}$, one needs $f_a \lesssim 3.0 \times 10^8$ 
GeV (correspondingly, $m_a \gtrsim 0.02$ eV). 
The more general parameter space in terms of $V_{\overline{126}}$ 
and $V_{16}$ is displayed in \fig{fig:HQPQ}.

In the post-inflationary PQ breaking scenario there is a main 
phenomenological issue given by the axion domain wall problem 
(recall that the model has $N_{\rm DW} = 12$, see \sect{sec:axionDW}).
A small breaking of the $\U(1)_{\rm PQ}$ via the Planck suppressed operator 
in \eq{eq:PQbreakPQquality} could remove the degeneracy among the 
vacua of the axion potential, effectively leading to the domain walls to 
decay before they dominate the energy density of the universe \cite{Sikivie:1982qv}. 
Parametrizing the PQ breaking via a bias term in the scalar potential 
\cite{Kawasaki:2014sqa,Harigaya:2018ooc}
\beq 
\label{eq:PQbias} 
\V_{\rm bias} = -2 \Xi V^4 \cos \( \frac{a}{V} + \delta \) \, , 
\eeq 
with $V = N_{\rm DW} f_a$, 
the energy density difference between two neighbouring minima, 
e.g.~in $a=0$ and $a=2\pi/N_{\rm DW}$, is (for $\sin \delta \approx 1$)
$\Delta \V = \V_{\rm bias} (2\pi/N_{\rm DW}) -  \V_{\rm bias} (0) 
\approx \Xi V^4 ( 1 
- \cos (2\pi / N_{\rm DW}) )$. 
A simple estimate of the 
decay time of the domain wall is obtained by equating the volume pressure 
$p_V \sim \Delta \V$ and the tension force $p_T 
\sim \sigma_{\rm DW} / t$ (with $\sigma_{\rm DW} \approx 9 m_a f_a^2$ 
denoting the wall tension \cite{Huang:1985tt,Hiramatsu:2012sc}). 
Hence,  
\beq 
t_{\rm decay} \approx \frac{\sigma_{\rm DW}}{\Xi V^4 ( 1 
- \cos (2\pi / N_{\rm DW}) ) } 
\approx 
5 \times 10^{-5} \ \text{s} \ 
\( \frac{10^{-50}}{\Xi} \)
\( \frac{12}{N_{\rm DW}} \)^4 
\( \frac{m_a}{0.02 \ \text{eV}} \)^{3}
\, . 
\eeq
Matching the bias parameter with the potential in \eq{eq:PQbreakPQquality} 
due to the Planck-suppressed operator $\phi_{16}^6 \phi^3_{\overline{126}}$, 
we obtain $\Xi \sim (V / M_{\rm Pl})^5 \approx 
5 \times 10^{-48} (0.02 \ \text{eV} /m_a)^5 (N_{\rm DW} / 12)^5$.
Hence, for $m_a$ in the ballpark of $0.02$ eV 
the domain walls safely decay before Big Bang nucleosynthesis 
(and also well before they start to dominate the energy density of the Universe). 
Note, however, that due to the $t_{\rm decay} \propto m_a^8$ dependence 
the bound saturates fast for larger $m_a$ (e.g.~$t_{\rm decay} \sim 10$ s 
for $m_a = 0.2$ eV). 

For $m_a \gtrsim 0.02$ eV, astrophysical constraints 
are also at play (see Sect.~4.7 in \cite{DiLuzio:2020wdo} for an updated summary). 
In particular, limits from red giants and white dwarfs cooling on the axion-electron coupling 
and the SN1987A bound on the axion-nucleon couplings are in the ballpark of 
$m_a \sim 0.02$ eV, depending however on electroweak vacuum parameters 
(cf.~\eqs{eq:axioncu}{eq:axioncde}). Their perturbativity 
domain will differ from those of the standard
DFSZ \cite{Zhitnitsky:1980tq,Dine:1981rt} 
axion model 
(as it does happen in PQ-extended left-right symmetric models \cite{Bertolini:2020hjc}). 
Hence, it is possible 
that astrophysical constraints could be relaxed to some extent compared to the 
DFSZ axion case, 
thus entering the region probed by IAXO in \fig{fig:axionphoton}.  
A detailed analysis of astrophysical constraints 
is beyond the scopes of the present work, but we note in 
passing that the axion-electron coupling 
$g_{ae} = c_e m_e / f_a$ (with $c_e$ given in \eq{eq:axioncde}), 
is of the size required to explain the 
so-called ``stellar cooling 
anomalies'' \cite{Giannotti:2017hny} for $m_a \approx 0.02$ eV and $c_e \approx 0.1$.  

Finally, it is interesting to note that recent simulations of the axion-string network 
suggest a much larger contribution to the total axion relic density than 
what traditionally thought \cite{Gorghetto:2018myk,Gorghetto:2020qws}. Rescaling the 
results of Ref.~\cite{Gorghetto:2020qws} 
for a generic domain wall number $N_{\rm DW}$, 
$m_a(N_{\rm DW}) = (Q (N_{\rm DW}) / Q (N_{\rm DW} =1))^{6/7} 
m_a(N_{\rm DW}=1)$ (with the function $Q$ given in Eq.~(36) of \cite{Gorghetto:2020qws}, 
after replacing $\xi_{\star} \log_{\star} \to N^2_{\rm DW} \xi_{\star} \log_{\star}$),  
one obtains 
a lower bound on the axion DM mass for $N_{\rm DW} = 12$ 
of about $m_a \gtrsim 7.5$ meV, 
which still neglects extra strings-domain walls 
contribution after the onset of axion oscillations.  
Hence, the requirement of the axion comprising the whole 
DM could potentially be compatible also with the high-quality axion mass window 
in \fig{fig:axionphoton}.

\section{Conclusions}
\label{sec:concl}

Obtaining an automatic $\U(1)_{\rm PQ}$ 
in GUTs 
is a longstanding problem, with remarkably 
few successful 
attempts.\footnote{The original work of Georgi et al.~\cite{Georgi:1981pu} 
based on SU(9) stands out as possibly the only successful one, 
that relies only on GUT dynamics. 
Ref.~\cite{Holman:1992us} proposed a SUSY $E_6 \times \U(1)'$ 
gauge model, where the $\U(1)_{\rm PQ}$ arises accidentally and 
it is protected against higher-dimensional operators. 
However, as observed in \cite{Dobrescu:1996jp}, 
the solution to the PQ quality problem is 
spoiled by soft SUSY breaking effects. 
Other approaches to the PQ quality problem based on composite 
dynamics such as \cite{Redi:2016esr}, 
can be made compatible with GUTs.} 
Any progress 
in that direction
would certainly make 
the connection between 
two motivated frameworks 
(the axion and grand unification) 
a more convincing one. 
Here, we have shown that the gauging the 
SU(3)$_{f}$ flavour group in SO(10) 
leads to an automatic $\U(1)_{\rm PQ}$ 
if the SO(10) Higgs representations are properly chosen.  
Moreover, the PQ symmetry is protected also beyond the renormalizable level, 
thus providing a possible way to tackle the PQ quality problem. 
In particular, the leading contributions to the axion potential 
arise from $d=9$ operator which, if Planck-suppressed, 
imply that the axion must be relatively heavy, $m_a \gtrsim 0.02$ eV. 
Disregarding instead the PQ quality problem in its standard formulation 
in terms of power-like effective operators, 
the axion can be as light as about $7 \times 10^{-8}$ eV, 
as implied by the upper bound on $f_a$ given by the seesaw scale. 
What emerges is an \emph{intermediate-scale} axion, that is quite different 
from the usual \emph{GUT-scale} axion, $m_a \lesssim 10^{-9}$ eV, 
obtained by dominantly breaking the PQ symmetry at the GUT scale. 
In the latter case, the $\U(1)_{\PQ}$ breaking SO(10) representations, 
i.e.~45, 54 or 210, have a trivial SO(10) center and thus the PQ symmetry 
does not arise automatically. From this perspective, the GUT-scale 
axion and in general $\SO(10) \times \U(1)_{\PQ}$ models in which the PQ breaking is 
related to the VEV of a complex 45, 54 or 210 
(even at intermediate mass scales, 
see e.g.~\cite{Reiss:1981nd,Lazarides:1981kz,Holman:1982tb,Altarelli:2013aqa,Ernst:2018bib,Boucenna:2018wjc})
appear to be theoretically less motivated, 
since it is 
more difficult to obtain an automatic $\U(1)_{\rm PQ}$ 
(compared to models in which the $\U(1)_{\rm PQ}$ is 
broken by SO(10) representations with non-trivial $B-L$ 
\cite{Mohapatra:1982tc,Chang:1987hz,Bajc:2005zf}).

Some of the considerations above extend as well to other GUT groups. 
Recent non-GUT constructions based on $\SU(\N)$ gauge dynamics 
\cite{DiLuzio:2017tjx,Ardu:2020qmo}, 
showed that it is possible to protect the PQ symmetry via the 
$\mathbb{Z}_\N$ center of $\SU(\N)$
up to operators of dimension $\N$. 
One could try to follow a similar path for GUTs. 
SU(5) is broken to the SM in one step, either via a 24 or a 75, 
both of which have however a trivial center 
(in fact, all SU(5) representations containing a SM-singlet direction have also zero quintality), thus making structurally difficult to get an automatic $\U(1)_{\rm PQ}$ 
in SU(5) where the PQ symmetry is broken via a complex adjoint \cite{Wise:1981ry,DiLuzio:2018gqe,FileviezPerez:2019ssf,FileviezPerez:2019fku}.  
Non-minimal GUTs such as SU(6) or $E_6$ are more promising in this respect, 
since they feature representations with SM-singlet directions 
(which can achieve large VEVs compared to the electroweak scale) 
and at the same time transform non-trivially under their respective 
$\mathbb{Z}_6$ and $\mathbb{Z}_3$ centers.  
Pati-Salam instead looks much alike SO(10).  
For instance, the models 
discussed in Refs.~\cite{DiLuzio:2020xgc,Saad:2017pqj} 
feature a PQ breaking representation that transforms trivially under the 
$\mathbb{Z}_4$ center of $\SU(4)_{\rm PS}$, and hence the $\U(1)_{\rm PQ}$ is not 
protected. Following a similar path as in the present work, 
an accidental Pati-Salam axion could arise by breaking the PQ symmetry 
with representations that have non-trivial $B-L$ quantum numbers (e.g.~a 4 and a 10) 
and transform non-trivially under flavour. 

Finally, 
it is remarkable that if flavour has to play a role for the $\U(1)_{\PQ}$ to arise accidentally, 
three generations are indeed quite special, since for example 
$n_g = 1,2,4$ would have not worked \cite{Chang:1984ip,Chang:1987hz}.  
So, ironically, $n_g = 3$ could have been responsible for making the CKM phase 
physical 
and at the same time washing out 
CP violation from strong interactions.

\begin{small}

\section*{Acknowledgments} 
I thank Marco Gorghetto, Federico Mescia, Michele Redi and Giovanni Villadoro 
for useful discussions. 
This work is supported by the Marie Sk\l{}odowska-Curie 
Individual Fellowship grant AXIONRUSH (GA 840791) 
and the Deutsche Forschungsgemeinschaft under Germany's Excellence Strategy 
- EXC 2121 Quantum Universe - 390833306.

\appendix

\section{Low-energy axion couplings}
\label{sec:loweaxion}

In this Appendix we provide a derivation of low-energy SO(10) axion couplings 
to SM matter fields,\footnote{See also 
Ref.~\cite{Ernst:2018bib} for a similar approach to SO(10) axion couplings. 
Note that the latter 
paper did not consider the specific SO(10) model 
studied here.} that requires the inclusion of electroweak VEVs 
for the identification of the physical axion field.  
\Table{tab:fieldsEW} summarizes the global (PQ) and gauge 
($B-L$ and $Y$) charges of the SO(10) sub-multiplets 
which host the axion as an angular component 
(this requires a complex scalar with a $Q = T^3_L + Y = 0$ 
neutral component), 
i.e. 
\beq 
\label{eq:LPhidec}
\Phi_i \supset \frac{V_{i}}{\sqrt{2}} e^{i \frac{a_{i}}{V_{i}}} \, , 
\eeq
with $\Phi = \{ \Delta^{\overline{126}}, \delta^{\overline{126}}, 
H^{16}_d, H^{10}_u, H^{10}_d, H^{\overline{126}}_u, H^{\overline{126}}_d, \Delta^{\overline{126}}_L \}$ spanning over the fields in \Table{tab:fieldsEW}. 

\begin{table}[t]
$$\begin{array}{c|c|c|c|c|c|c|c|c}
& \Delta^{\overline{126}} 
& \delta^{16} 
& H^{16}_d 
& H^{10}_u
& H^{10}_d
& H^{\overline{126}}_u
& H^{\overline{126}}_d
& \Delta^{\overline{126}}_L 
\\ \hline
\PQ & -2 & -1 & -1 & -2 & -2 & -2 & -2 & -2 \\
B-L & -2 & 1 & -1 & 0 & 0 & 0 & 0 & 2 \\
Y & 0 & 0 & -1/2 & 1/2 & -1/2 & 1/2 & -1/2 & 1 \\
\end{array}$$
\caption{Global (PQ) and local ($B-L$ and $Y$) 
charges of the SO(10) 
sub-multiplets hosting the physical axion field.}
\label{tab:fieldsEW}
\end{table}%

The low-energy PQ symmetry, with charge $q$, is a linear combination 
of the UV PQ charges and two broken Cartan generators, 
which can be chosen as $B-L$ and $Y$, namely 
\beq 
\label{eq:generalUVIRPQ2}
q = c_1 \PQ + c_2 (B-L) + c_3 Y \, .  
\eeq
It can be shown, analogously to the cases discussed in \sect{sec:physaxion}, 
that in order to match $\U(1)_{\PQ}$ anomalies in terms of UV and IR charges, 
$c_1 = 1$, which we henceforth assume in the following. 
Given the PQ current $J_\mu^{\rm PQ} = \sum_i q_{i} V_{i} \partial_\mu a_{i}$
The canonical axion field is defined as 
\beq 
\label{eq:axiondef2}
a = \frac{1}{V} \sum_i q_{i} V_{i} a_{i} \, , \qquad 
V^2 = \sum_i q^2_{i} V^2_{i} \, , 
\eeq
so that $J_\mu^{\rm PQ} = V \partial_\mu a$ and, 
compatibly with the Goldstone theorem, 
$\langle 0 | J_\mu^{\rm PQ} | a \rangle = i V p_\mu$.  
Under a PQ transformation $a_{i} \to a_{i} 
+ \alpha q_{i} V_{i}$ 
the axion field transforms as $a \to a + \alpha V$. 
Inverting the orthogonal transformation in \eq{eq:axiondef2}, 
one readily 
obtains the 
projection of the angular modes on the axion field: 
\beq 
\label{eq:aproj2}
a_{i} \to q_{i} V_{i} \frac{a}{V} \, .  
\eeq 
To determine the $q$ charges we proceed as follows. 
First, we require the orthogonality between the axion current and the 
gauge currents $J_{B-L} = \sum_i (B-L)_{i} V_{i} \partial_\mu a_{i}$ 
and $J_Y = \sum_i Y_{i} V_{i} \partial_\mu a_{i}$ (to avoid kinetic mixings of the axion field 
with massive gauge bosons). This yields, respectively: 
\begin{align}
-2 q_{\overline{126}} V^2_{\overline{126}} 
+ q_{16} V^2_{16} 
- q_{H_d^{16}} (v^d_{16})^2 &= 0  \, ,  \\
- \frac{1}{2} q_{H_d^{16}} (v^d_{16})^2 
+ \frac{1}{2} q_{H_u^{10}} (v^u_{10})^2 
- \frac{1}{2} q_{H_d^{10}} (v^d_{10})^2
+ \frac{1}{2} q_{H_u^{\overline{126}}} (v^u_{\overline{126}})^2 
- \frac{1}{2} q_{H_d^{\overline{126}}} (v^d_{\overline{126}})^2 &= 0 \, .   
\end{align}
Second, by decomposing the invariants with non-trivial global re-phasings 
in the SO(10) scalar potential (cf.~\eqs{eq:V3}{eq:V3}):\footnote{In the following, 
we set to zero
the $\SU(2)_L$ triplet VEV, $v_{\Delta^{\overline{126}}_L}$. 
Keeping the latter, axion couplings to SM charged fermions 
receive safely negligible 
corrections of the order of 
$v^2_{\Delta^{\overline{126}}_L} / v^2 \ll 1$ (with $v$ denoting the electroweak scale) \cite{BDLN}.} 
\begin{align}
&\phi^2_{16} \phi^{\star}_{10} \supset H_d^{16} \delta^{16} (H_d^{10})^\star \, , \\
&\phi^2_{10} \phi^{\star 2}_{\overline{126}} 
\supset (H_u^{10})^2 (H_u^{\overline{126}})^{2\star}\, , (H_d^{10})^2 (H_d^{\overline{126}})^{2\star} \, , 
\end{align}
one obtains the 
following extra constraints on $q$ charges (recall \eq{eq:LPhidec} and \eq{eq:aproj2}): 
\begin{align}
q_{H_d^{16}} + q_{16} - q_{H_d^{10}} &= 0 
\, , \\
q_{H_u^{10}} -  q_{H_u^{\overline{126}}} = 
q_{H_d^{10}} -  q_{H_d^{\overline{126}}} &= 0 \, ,
\end{align}
from which we see that the (electroweak) $q$ charges of $\phi_{10}$ and 
$\phi_{\overline{126}}$ are aligned. 

To close the system of linear equations, in order to extract the $q$ charges, 
it is necessary to include also the matching between UV and IR PQ charges 
in \eq{eq:generalUVIRPQ2} for all the scalar fields.   
Here, we present the result in the simplifying limit $v_{H^{16}_d} \to 0$  
(although for the final expressions 
of axion couplings we will keep $v_{H^{16}_d}\neq0$), 
that is
\begin{align}
c_2 &= \frac{V^2_{16} - 4 V^2_{\overline{126}}}{V^2_{16} + 4 V^2_{\overline{126}}} \, , \\
c_3 &= 4 \frac{v^2_u - v^2_d}{v^2} \, , \\
q_{\overline{126}} &= - \frac{4 V^2_{16}}{V^2_{16} + 4V^2_{\overline{126}}} \, , \\
q_{16} &= - \frac{8 V^2_{\overline{126}}}{V^2_{16} + 4V^2_{\overline{126}}} \, , \\
q_{H_d^{16}} &= \frac{8 V^2_{\overline{126}} (v^2_d - v^2_u) - 
4 V^2_{16} v^2_u}{(V^2_{16} + 4V^2_{\overline{126}})v^2}  \, , \\
q_{H_u^{10}} &= q_{H_u^{\overline{126}}} = -4 \frac{v^2_d}{v^2}  \, , \\
q_{H_d^{10}} &= q_{H_d^{\overline{126}}} = -4 \frac{v^2_u}{v^2} \, , \\
\label{eq:V2expr}
V^2 &= 16 \( \frac{V^2_{16}V^2_{\overline{126}}}{V^2_{16} + 4 V^2_{\overline{126}}} 
+ \frac{v_u^2v_d^2}{v^2} \) \, .
\end{align}
where we defined 
$v_u^2 = v_{H_u^{10}}^2 + v_{H_u^{\overline{126}}}^2$, 
$v_d^2 = v_{H_d^{10}}^2 + v_{H_d^{\overline{126}}}^2$ 
and  
$v^2 = v^2_u + v^2_d$.

To compute low-energy axion couplings to SM charged fermions, 
we decompose $\psi_{16}\psi_{16}\phi_{10} \supset q u^c H^{10}_u + q d^c H^{10}_d 
+ \ldots$. The axion is removed from the Yukawa interaction via 
a family universal 
axion-dependent transformation (in Weyl notation and suppressing generation indices)
\begin{align}
&u \to e^{-i q_{H_u^{10}} \frac{a}{2V}}  u \, , 
&u^c \to e^{-i q_{H_u^{10}}\frac{a}{2V}}  u^c \, , \\
&d \to e^{-i q_{H_d^{10}} \frac{a}{2V}} d \, , 
&d^c \to e^{-i q_{H_d^{10}} \frac{a}{2V}} d^c \, , 
\end{align}
(equivalently, in Dirac notation, 
$u \to e^{i q_{H_u^{10}} \gamma_5 \frac{a}{2V} } u$, etc.), 
and analogously for charged leptons. 
The same rotation also removes the axion from the other Yukawa interaction, 
$\psi_{16}\psi_{16}\phi^\star_{\overline{126}}$, due to the 
alignment of the (electroweak) PQ charges of 
$\phi_{10}$ and $\phi_{\overline{126}}$. 

The SM chiral fermion transformation above generate the anomalous terms 
\beq 
\delta \mathscr{L} = 
\frac{\alpha_s N}{4 \pi} \frac{a}{V} G \tilde G + 
\frac{\alpha E}{4 \pi} \frac{a}{V} F \tilde F \, ,
\eeq 
with the anomaly factors 
\begin{align}
N &= - n_g T(3) \(2 \frac{q_{H_u^{10}}}{2} + 2 \frac{q_{H_d^{10}}}{2} \) = 6 \, , \\
E &= - n_g \(q_{H_u^{10}} 3 (2/3)^2 + q_{H_d^{10}} 3 (-1/3)^2 
+ q_{H_d^{10}} (-1)^2 \) = 16 \, ,  
\end{align} 
using $n_g = 3$, $T(3) = 1/2$, etc.  
So, in particular, upon choosing the standard normalization of the 
axion-gluon coupling in terms of the Lagrangian term 
\beq 
\frac{\alpha_s}{8 \pi} \frac{a}{f_a} G \tilde G \, ,
\eeq
the axion decay constant reads (using \eq{eq:V2expr} with the sign fixed as in \eq{eq:axiondc} for the case of no electroweak VEVs)
\beq 
\label{eq:deffaV}
f_a = \frac{V}{2N} = - \frac{1}{3} \sqrt{\frac{V^2_{16}V^2_{\overline{126}}}{V^2_{16} + 4 V^2_{\overline{126}}} 
+ \frac{v_u^2v_d^2}{v^2} } \, , 
\eeq
while $E/N = 8/3$, which enters the axion-photon coupling \cite{diCortona:2015ldu}
\beq 
\label{eq:axionph}
C_{a\gamma} =  E/N - 1.92(4) \, ,
\eeq 
defined in terms of the Lagrangian term $\mathscr{L} \supset \frac{\alpha}{8\pi f_a} C_{a\gamma} a F \tilde F$.  

On the other hand, the variation of the fermion kinetic terms yields
(in Dirac notation)
\begin{align}
\delta(\bar u i \slashed{\partial} u) 
&= - q_{H_u^{10}} \frac{\partial_\mu a}{2 V} \bar u \gamma^\mu \gamma_5 u \, , \\
\delta(\bar d i \slashed{\partial} d) 
&= - q_{H_d^{10}} \frac{\partial_\mu a}{2 V} \bar d \gamma^\mu \gamma_5 d \, , \\
\delta(\bar e i \slashed{\partial} e) 
&= - q_{H_d^{10}} \frac{\partial_\mu a}{2 V} \bar e \gamma^\mu \gamma_5 e \, . 
\end{align}
Hence, defining the axion coupling to SM fermions via 
$\mathscr{L} \supset \frac{\partial_\mu a}{2 f_a} c_f \bar f \gamma_\mu \gamma_5 f$, 
and replacing $f_a = V / (2N)$, we find the (family universal) couplings
\begin{align}
\label{eq:axioncu}
c_u &= 
- \frac{q_{H_u^{10}}}{2N} =
\frac{1}{3} \frac{V^2_{16} (v_d^2 + v_{H^{16}_d}^2) 
+ 2 V^2_{\overline{126}} (2 v_d^2 + v_{H^{16}_d}^2) 
+ v_d^2 v_{H^{16}_d}^2 }{
(V^2_{16} + 4 V^2_{\overline{126}}) (v^2 + v_{H^{16}_d}^2) 
+ v^2 v_{H^{16}_d}^2
}
\approx  \frac{v^2_d}{3v^2}  \, , \\
\label{eq:axioncde}
c_d &= c_e 
=  - \frac{q_{H_u^{10}}}{2N} = 
\frac{1}{3} \frac{V^2_{16} v_u^2  
+ 2 V^2_{\overline{126}} (2 v_u^2 + v_{H^{16}_d}^2) 
+ v_u^2 v_{H^{16}_d}^2 }{
(V^2_{16} + 4 V^2_{\overline{126}}) (v^2 + v_{H^{16}_d}^2) 
+ v^2 v_{H^{16}_d}^2
}
\approx \frac{v^2_u}{3v^2}  \, , 
\end{align}
where in the last step we provided 
simplified expressions in the $v_{H^{16}_d} \to 0$ limit.

\bibliographystyle{utphys.bst}
\bibliography{bibliography}

\end{small}

\end{document}